\documentclass[12pt, a4paper, leqno]{article}
\def\rr{{\hbox{I{\kern -0.22em}R}}}

\newcommand{\singlespacing}{\renewcommand{\baselinestretch}{1.1} \normalsize}
\newcommand{\onehalfspacing}{\renewcommand{\baselinestretch}{1.3} \normalsize}

\newcommand{\eof}{\hfill {\it Q.E.D.} \vspace*{0.3cm}}

\newcounter{neweqn}
\newcounter{newthm}
\newcounter{newlemma}
\newcounter{newcor}
\newcounter{newpro}
\newcounter{newDef}
\newcounter{newrem}
\newcounter{neweg}
\newcounter{newsec}

\newcommand{\newsection}[3]{\begin{center}
\refstepcounter{newsec} \label{#2} \thenewsec. {\rm #1}
\end{center}
 \baselineskip#3
\setcounter{neweqn}{0} \setcounter{equation}{0}
\setcounter{newthm}{0} \setcounter{thm}{0}
\setcounter{newlemma}{0} \setcounter{lemma}{0}
\setcounter{newcor}{0} \setcounter{cor}{0}
\setcounter{newpro}{0} \setcounter{pro}{0}
\setcounter{newrem}{0} \setcounter{remark}{0}
\setcounter{newDef}{0} \setcounter{Def}{0}
\setcounter{neweg}{0} \setcounter{eg}{0}}

\newcommand{\bq}[1]{\begin{equation} \refstepcounter{neweqn} \refstepcounter{equation}\label{#1}}

\newcounter{thm}
\newcounter{lemma}
\newcounter{cor}
\newcounter{pro}
\newcounter{Def}
\newcounter{remark}
\newcounter{eg}

\newcommand{\bthm}[1]{\vspace{2mm} \refstepcounter{newthm}
T\footnotesize HEOREM \normalsize \thethm: \itshape
\refstepcounter{thm} \label{#1}}
\newcommand{\blemma}[1]{\vspace{2mm} \refstepcounter{newlemma}
L\footnotesize EMMA \normalsize \thelemma:
\itshape  \refstepcounter{lemma} \label{#1}}

\newcommand{\bpro}[1]{\vspace{2mm} \refstepcounter{newpro}
P\footnotesize ROPOSITION \normalsize \thepro:
\itshape  \refstepcounter{pro} \label{#1}}
\newcommand{\bDef}[1]{\vspace{2mm} \refstepcounter{newDef}
D\footnotesize EFINITION \normalsize \theDef:
\refstepcounter{Def} \label{#1}}
\newcommand{\brem}[1]{\vspace{2mm} \refstepcounter{newrem}
R\footnotesize EMARK \normalsize \theremark:
\refstepcounter{remark} \label{#1}}
\newcommand{\beg}[1]{\vspace{2mm} \refstepcounter{neweg}
E\footnotesize XAMPLE \normalsize \theeg:
\refstepcounter{eg} \label{#1}}

\newcommand{\ethm}{\vspace{2mm} \upshape}
\newcommand{\elemma}{\vspace{2mm} \upshape}

\newcommand{\epro}{\vspace{2mm} \upshape}
\newcommand{\eDef}{\vspace{2mm} }
\newcommand{\erem}{\vspace{2mm} }
\newcommand{\eeg}{\vspace{2mm} }

\renewcommand{\thethm}{{\rm \thenewsec.\thenewthm}}
\renewcommand{\thelemma}{{\rm \thenewsec.\thenewlemma}}

\renewcommand{\thepro}{{\rm \thenewsec.\thenewpro}}
\renewcommand{\theDef}{{\rm \thenewsec.\thenewDef}}
\renewcommand{\theremark}{{\rm \thenewsec.\thenewrem}}
\renewcommand{\theeg}{{\rm \thenewsec.\theneweg}}

\newcommand{\eq}{\end{equation}}

\newcommand{\bd}{\begin{displaymath}}

\newcommand{\ed}{\end{displaymath}}
\newcommand{\ba}{\bd\begin{array}{rl}}
\newcommand{\ea}{\end{array}\ed}

\newcommand{\Ga}{\Gamma}

\newcommand{\si}{\sigma}

\newcommand{\cds}{\cdots}

\newcommand{\cF}{{\cal F}}

\newcommand{\su}{\sum_{j=1}^m}

\newcommand{\as}{\mbox{{\rm a.s.}}}

\newcommand{\refs}[1]{(\ref{#1})}

\usepackage{amsmath}
\usepackage{array,tabularx}
\usepackage{epsfig}
\usepackage{pst-node,pstricks}
\usepackage{supertabular}
\usepackage{longtable}
\usepackage{multicol}
\usepackage{subfigure}
\usepackage{latexsym}
\usepackage{graphicx}
\usepackage{amsmath}

\begin{document}



\onehalfspacing

\begin{center}
THE PREMIUM OF DYNAMIC TRADING\footnote{This research was supported by
the RGC Earmarked Grants
CUHK 4175/03E, CUHK418605, and Croucher Senior Research Fellowship.}

\vspace{5mm}

{\sc By}
{\sc Chun Hung Chiu}\footnote{Business Section, Institute of Textile and Clothing, The Hong Kong Polytechnic University, Hung Hom, Kowloon, Hong Kong.
E-mail: $<$tcchiu@polyu.edu.hk$>$.}
AND
{\sc Xun Yu Zhou}\footnote{
Nomura Centre for Mathematical Finance, and Oxford--Man Institute of Quantitative Finance,  University of Oxford,
24--29 St Giles, Oxford OX1 3LB, and Department of Systems Engineering and Engineering Management,
The Chinese University of Hong Kong, Shatin, Hong Kong.
E-mail: $<$zhouxy@maths.ox.ac.uk$>$;
Tel.: +44(0)1865-280614; Fax: +44(0)1865-270515.}



\newpage

\begin{center}
{\large THE PREMIUM OF DYNAMIC TRADING}
\end{center}

\bigskip

\singlespacing
\parbox{5.3in}{ \hspace{5mm} \footnotesize
It is well established
that in a market with inclusion of a risk-free asset
the {single-period} mean--variance efficient frontier is a straight line tangent to
the risky region, a fact that is the very foundation of the classical CAPM. In this paper, it is shown that
in a continuous-time market where the risky prices are described by
It\^o's processes and the investment opportunity set is
deterministic (albeit time-varying), {any} efficient
portfolio must involve allocation to the risk-free asset at {any} time.
As a result,
the dynamic mean--variance efficient frontier, though
still a straight line, is {strictly above} the entire risky region.
This in turn suggests a positive premium, in terms of the Sharpe ratio of the efficient
frontier, arising
from the dynamic trading. Another implication is that
the inclusion of a risk-free asset boosts the
Sharpe ratio of the efficient frontier,
which again contrasts sharply with the single-period case.

\normalsize}
\vspace{5mm}

\parbox{5.3in}{\hspace{5mm} \footnotesize
K\scriptsize EYWORDS: \footnotesize Continuous time, portfolio selection,
mean--variance efficiency, Sharpe ratio}
\normalsize
\end{center}



\onehalfspacing

\bigskip

\newsection{INTRODUCTION}{s1}{0.28in}

\begin{figure}[h]
  \begin{center}
    \includegraphics[scale=0.65]{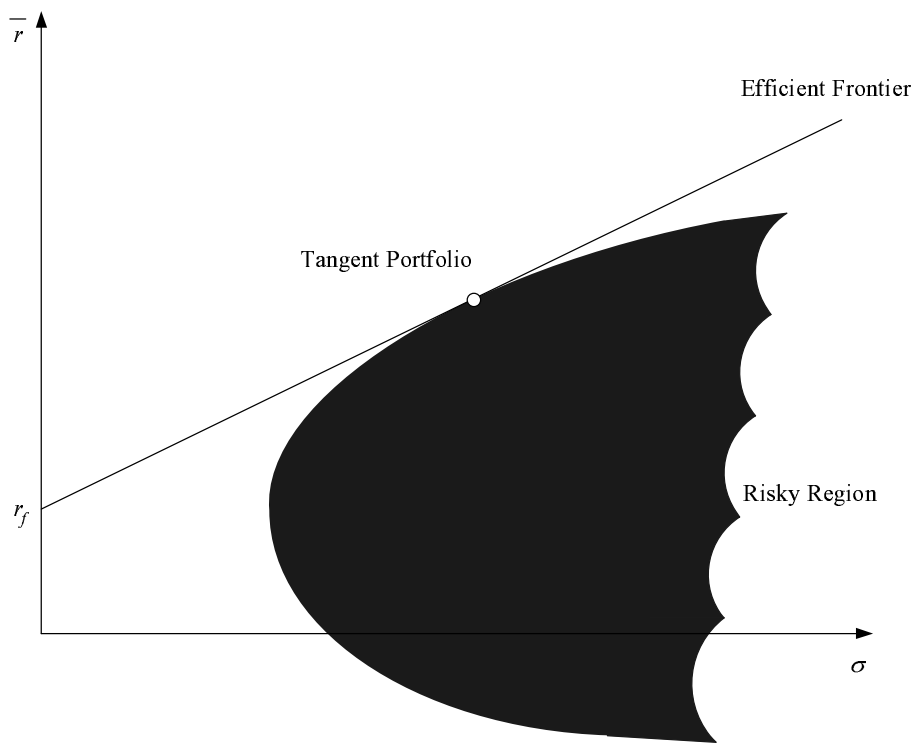}
    \caption{Efficient Frontier and Risky Region: Single-Period Case}\label{fig1a}
    \end{center}
\end{figure}

Given a single investment period and a market where
there are a number of basic assets including a
risk-free one,
Markowitz's classical
mean--variance theory [Markowitz (1952, 1987), Merton (1972)]
stipulates that, when the risk-free asset is available,  the efficient frontier
is a straight line tangent to the risky hyperbola\footnote{Any portfolio
using the basic assets can be mapped onto
the {\it mean--standard {\red deviation} diagram},
a two-dimensional diagram where the mean and standard deviation of
the portfolio rate of return are used
as the vertical and horizontal axes, respectively. The {\it risky hyperbola}
is
the left boundary
of the region -- the risky region {\red (see Section 2 for the precise definition of the risky region)} -- on the diagram representing all the possible portfolios
formed from the basic {risky} assets.}; see Figure \ref{fig1a}.
Several fundamentally important conclusions have been drawn from
this fact: 1) There is at least one portfolio (called a {\it
tangent portfolio} or {\it tangent fund}), composed of the
basic {risky} assets {only}, that is efficient;
2) If one uses Sharpe ratio to measure the reward-to-risk\footnote{Sharpe ratio is the ratio between the excess rate of return
(over the risky-free rate) and the standard
deviation of the return rate of a portfolio.
Here the risk-free rate serves as
a reference point or benchmark only; a portfolio's
Sharpe ratio is defined regardless of
whether or not a risk-free asset is available for inclusion
when constructing the portfolio.},
then
the inclusion of a risk-free asset does {not}
increase the {highest} Sharpe ratio s/he could possibly achieve with the basic risky assets;
3) Any efficiency-seeker needs only
to invest on the risk-free asset and the tangent fund
with a suitable proportion in accordance with her/his risk taste; this observation
is called the mutual fund theorem [Tobin (1958)]; and 4) At demand--supply equilibrium
the market portfolio is nothing else than the tangent portfolio, and
the expected excess rate of return of any individual asset
is linearly related to its beta -- a.k.a. the Sharpe--Lintner--Mossin
capital
asset pricing model [abbreviated as CAPM; see Sharpe (1964), Lintner (1965), and Mossin (1966)].

So, Figure \ref{fig1a} visualizes some of the most important
results in modern portfolio theory and asset pricing theory. Questions
we would like to answer in this paper are: what is the corresponding
figure for a {\it dynamic} market which allows {\it continuous}
trading?\footnote{Although in this paper we work within
the continuous-time framework, all the discussions and results carry over
to the dynamic, discrete-time setting.} What are the
implications if the figure in the continuous-time setting becomes different?

To address these questions, one needs to solve the dynamic Markowitz
problem first.
Happily, the dynamic extension of
the Markowitz model, especially in continuous time, has been studied
extensively in recent years; see, e.g.,
Richardson (1989), Bajeux-Besnainous and Portait (1998), Li and Ng (2000), 
Zhou and Li (2000), 
and Lim (2004). 

In many of the above-cited works on dynamic Markowitz's problems,
explicit, analytic forms of
efficient portfolios have been obtained. In particular, if the
investment opportunity set is deterministic (though possibly time-varying)
and portfolios are unconstrained, then the efficient frontier
 is shown by Bajeux-Besnainous and Portait (1998) and
Zhou and Li (2000) to remain a straight line in a continuous-time
market\footnote{The efficient frontier
in the continuous-time setting is plotted on the mean--standard deviation diagram
at the {\it end} of the investment horizon.
We are interested in the terminal time only because the two criteria of a
mean--variance
model -- mean and variance, that is -- concern only the terminal payoff.}.
Based on this result together with the explicitly derived
efficient portfolios, we are going to show in this paper that the
efficient frontier is, indeed,  {\it strictly above} the risky region, as indicated
by Figure 2.

\begin{figure}[h]
\begin{center}
    \includegraphics[scale=0.65]{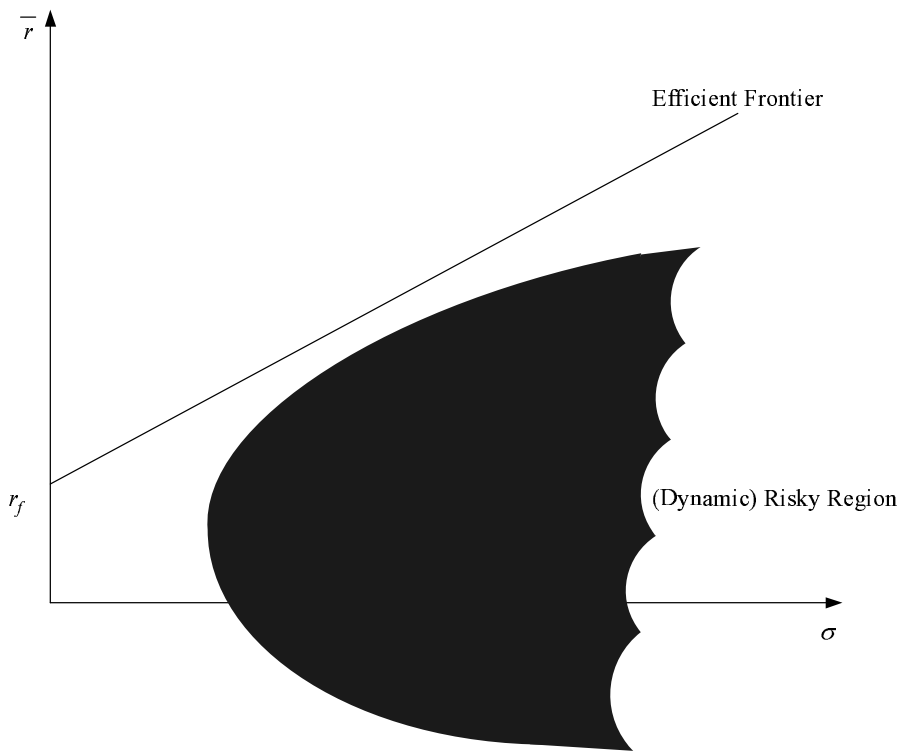}
    \caption{Efficient Frontier and Risky Region: Continuous-Time Case}\label{fig1b}
    \end{center}
\end{figure}

Notice that the risky region (the dark area) in Figure 2 represents all the
portfolios that could {\it continuously} rebalance among basic risky assets;
therefore
it is a much {\it expanded} region than the one corresponding to {\it static} (buy-and-hold) portfolios involving no transactions between the initial and terminal
times.
In Bajeux-Besnainous and Portait (1998), p.83,  a figure is
depicted showing that the dynamic efficient frontier is strictly
above a risky region. The figure there and Figure 2 here may
appear similar at the first glance;
yet a fundamental difference between the two
is that the risky region in Bajeux-Besnainous and Portait (1998)
is the one spanned by {\it buy-and-hold}
portfolios involving basic risky assets (hereafter referred to as the {\it buy-and-hold
risky region} to be distinguished from the {\it dynamic risky region} we are dealing with
in this paper).
For reasons explained above,
our (dynamic) risky region is much larger than the buy-and-hold risky region for the same
continuous-time market (the latter is {irrelevant} to the dynamic
economy anyway);
therefore our result is more powerful and more relevant to the dynamic
setting. Indeed, the figure in Bajeux-Besnainous and Portait (1998)
is nothing else than an (almost trivial) statement that a dynamic
efficient portfolio is strictly better than any buy-and-hold
portfolio, whereas our figure implies that
a dynamic
efficient portfolio is {\it strictly} better than {\it any} portfolio
that is allowed to continuously switch among risky assets\footnote{Only in
one special case, i.e., when there is merely one risky asset, do the
two figures coincide.}.

Figure 2
reveals a major and surprising departure of a dynamic economy
from a static one.
Immediate are the following consequences:
1) No portfolio consisting of only risky assets could be efficient. In other
words, any efficient portfolio must invest in the risk-free asset;
2) The efficient frontier line is pushed away from the (dynamic) risky region as
a result of the availability of dynamic trading. We call this
enhancement on the Sharpe ratio the
{\it premium of dynamic trading};
3) The inclusion of the risk-free asset in one's portfolio indeed
{\it strictly increases} the best Sharpe ratio achievable
compared with the case when s/he only has risky assets at disposal;
4) Efficient portfolios are no longer simple convex combinations of the risk-free
asset and a risky fund; so the mutual fund theorem (in the conventional
sense) fails\footnote{In
Bajeux-Besnainous and Portait (1998) a what is termed {\it strong separation}
is derived, which asserts that buy-and-hold strategies involving
the bond and an appropriate dynamic strategy describe the entire
efficient frontier. However, it is not mentioned there whether or not -- the
answer is no by virtue of {\it our} results -- {\it that} particular dynamic strategy is a
pure risky strategy as with a single-period model.};
5) The market portfolio is no longer mean--variance efficient even under
the supply-demand equilibrium, if the market portfolio is defined
analogous to that in the single-period setting\footnote{In the single-period
case the market portfolio is the tangent portfolio, which does {\it not} involve
allocation to a bond. It is the summation of all the basic {\it risky}
assets being circulated in the market.}
. The CAPM in the present
setting needs to be
studied more carefully.

What is the cause of such a drastic change in the dynamic
setting? An immediate answer might be that, there are much more
portfolios to {\red choose} from because of the possibility of dynamic trading,
and hence the admissible region\footnote{The admissible region is
the region on the mean--standard deviation
diagram representing all the possible portfolios involving {\it both} risky and risk-free assets.
The risky region is a strict subset of the admissible region.}
is much expanded than its
single-period counterpart. While this is indeed a good point, it
has yet to explain the puzzled phenomenon wholly: Why {\red is} the efficient
frontier completely pushed {\it away} from the dynamic risky region rather than, say, the whole admissible region is expanded whereas the frontier still
touches the risky hyperbola? Why {\red is there } a boost of Sharpe ratio when
the risk-free asset is included?

The true answer lies, very subtly, in the way
when the risk-free asset and a given risky asset combines to form a portfolio.
In the single-period case, portfolios using the two assets generate a
{\it straight line} connecting these two assets on the diagram.
In the continuous-time setting, because of the possibility
of continuously adjusting the weights between the two assets,
the resulting portfolios form
a solid, two-dimensional {\it region} with an upward curve as its upper boundary. It is these
infinitely many upward curves (one for every risky asset)
that eventually push the efficient frontier
away from the risky region. We defer more detailed discussions on this
to Section 4.

In deriving the aforementioned separation result, some
properties of a dynamic efficient policy will be revealed, which
are interesting in their own rights. These
properties, unique to the dynamic setting, include that
any efficient wealth process is capped by a deterministic constant, any non-trivial
efficient strategy must be exposed to risky assets at any time, and any efficient portfolio must invest in the risk-free asset at {\it any} time.

The remainder of the paper is organized as follows. In Section 2,
the market under consideration is described and the continuous--time
mean--variance formulation is presented.
In Section 3, some known results on the
mean--variance efficient portfolios and
frontier are highlighted. 
Section 4
is devoted to a
detailed study on the strict separation between the frontier and
the dynamic risky region, introducing the term of the premium of dynamic trading.
Section 5 offers economical explanations on the premium.
Section 6 concludes the paper. All the proofs are deferred to an
appendix.

\newsection{A CONTINUOUS-TIME MEAN--VARIANCE MARKET}{s2}{0.28in}
Throughout this paper $(\Omega,\cF,P,\{\cF_t\}_{t\geq0})$ is a
fixed filtered complete probability space on which defined a
standard $\cF_t$-adapted $m$-dimensional Brownian motion
$\{W(t), t\geq 0\}$ with
$W(t)\equiv(W^1(t),\cdots,W^m(t))'$ and $W(0)=0$.
It is assumed that $\cF_t=\si\{W(s):s\le t\}$ argumented by all the
$P$-null sets.
For a fixed $T>0$, we denote by
$L^2_\cF(0,T;\Re^d)$ the set of all $\Re^d$-valued, $\cF_t$-progressively measurable
stochastic
processes $f(\cdot)$ with
$E\int_0^T|f(t)|^2dt<+\infty$. Also,
we use $M'$ to denote
the transpose of any vector or matrix $M$, and $\sigma_\xi$ to denote the
standard deviation of a random variable $\xi$.

Suppose there is a capital market
in which $m+1$ {\it basic assets}
are being traded continuously. One of the assets
is a risk-free bond whose {value process} $S_0(t)$ is subject to the following
ordinary differential equation:
\bq{equ001}
\left\{\begin{array}{l}
dS_0(t)=r(t)S_0(t)dt,\;\; t\geq0,\\
S_0(0)=s_0>0,
\end{array}\right.
\end{equation}
where $r(t)>0$ is the {interest rate}.
The other $m$ assets are risky {stocks}
whose price processes $S_1(t),\cds, S_m(t)$ satisfy the following stochastic
differential equation (SDE):
\bq{equ002}
\left\{\begin{array}{l}
dS_i(t)=S_i(t)\big[\mu_i(t)dt+\sum_{j=1}^m\si_{ij}(t)dW^j(t)\big],\;\;
t\geq0,\\
S_i(0)=s_i>0,\;\; i=1,2,\cdots,m,
\end{array}\right.
\end{equation}
where $\mu_i(t)>0$ is the {appreciation rate}, and
$\si_{ij}(t)$
the {volatility} or dispersion rate of the stocks.
Here the investment opportunity set $\{r(\cdot),\mu_i(\cdot),\si_{ij}(\cdot),\;i,j=1,2,\cdots,m\}$
is deterministic (yet time-varying).

Define the excess rate of return vector
$B(t):=(\mu_1(t)-r(t), \ldots, \mu_m(t)-r(t))'$ and
the {covariance matrix}
$\si(t):=(\si_{ij}(t))_{m\times m}$.
We assume that $B(t)$ is a continuous function of $t$, and
\bq{non-degeneracy} \si(t)\si(t)'\geq \delta I,\;\;\; \forall
t\geq0,
\end{equation}
for some $\delta>0$.
These conditions ensure the market to be arbitrage-free and complete.

Consider an agent, with an initial endowment $x_0>0$ and an investment horizon $[0,T]$,
whose total wealth at time $t\ge0$ is denoted
by $x(t)$. Assume that the trading of shares is self-financed and
takes place continuously, and that transaction cost and consumptions are not considered. Then $x(\cdot)$ satisfies
[see, e.g., Karatzas and Shreve (1999)] 
\bq{system2}
\left\{\begin{array}{l}
dx(t)=[r(t)x(t)+B(t)'\pi(t)]dt+\pi(t)'\si(t)dW(t),\\
x(0)=x_0,
\end{array}\right.
\end{equation}
where $\pi(t):=(\pi_1 (t),\cdots,\pi_m(t))'$, with
$\pi_i(t),\; i=0,1,2\cds,m,$
denoting the total market value of the agent's wealth in the $i$-th
asset (in particular, $\pi_0(t)$ is the wealth invested in the bond
at $t$).
We call
$\pi(\cdot)$ a {\it portfolio} of the agent, which is a
stochastic process.
A portfolio $\pi(\cdot)$ is said to be {\it admissible} 
if
$\pi(\cdot) \in {L}_\mathcal{F}^2(0,T;\Re^m)$ 
and the SDE (\ref{system2}) has a unique solution
$x(\cdot)$ corresponding to $\pi(\cdot)$. In this case, we refer to
$(x(\cdot), \, \pi(\cdot))$ as an {\it admissible} ({\it wealth--portfolio})
{\it pair}. 
An admissible portfolio is also interchangeably referred to
as an (admissible) {\it asset}.

{\rm With an admissible portfolio $\pi(\cdot)$ the corresponding wealth
trajectory, $x(\cdot)$, is completely determined via the SDE (\ref{system2}). As a result,
$\pi_0(\cdot)$, the allocation to the bond,
 is derived as $\pi_0(t)=x(t)-\sum_{i=1}^m\pi_i(t)$.
This is the reason why we should not include $\pi_0(\cdot)$ in defining
a portfolio $\pi(\cdot)$.} {\red An admissible portfolio with $P\{\pi_0(t)=0\}=1$, a.e.$t\in [0,T]$ is called a {\it pure risky portfolio}.}

As with the single-period case a mean--standard {\red deviation} diagram
(hereafter referred to as {\it the diagram}) is
a two-dimensional diagram where the mean and standard deviation are used
as vertical and horizontal axes, respectively.
For any {admissible} wealth--portfolio
pair satisfying (\ref{system2}),
define ${R}(t): = (x(t) - x_0)/x_0$, i.e., the corresponding
return rate at $t$.
{\rm 
The set of all points $\left(\sigma_{R(T)}, E[R(T)]\right)$
on the diagram, where $R(T)$ is the return rate of an admissible
portfolio at $T$, is called the {\it admissible region}.
} {\red The subset of the admissible region corresponding to all the pure risky portfolios is called the {\it (dynamic) risky region}.}

The agent's objective is to find an admissible portfolio $\pi(\cdot)$, among all such
admissible portfolios that their
expected terminal wealth $E x(T) = z$, where $z \in \Re$ is given {\it a
priori},
so that
the risk measured by the variance of the terminal wealth
\bq{var0}
{\rm Var}\; x(T) := E[x(T) - Ex(T)]^2 \equiv E[x(T) - z]^2
\end{equation}
is minimized.
Geometrically, the problem is to
locate the left boundary of the admissible region.
Mathematically, we have the following
formulation.

\bDef{def002}
{\rm Fix the initial wealth $x_0$ and
the terminal time $T$.
The mean--variance portfolio selection
problem is formulated as a
constrained stochastic optimization problem parameterized by
$z\geq x_0 e^{ \int_0^T r(t)dt}$:
\bq{equ007}
\begin{array}{ll}
\mbox{minimize} &J_{{\rm MV}}(\pi(\cdot)) := E [x(T) - z]^2, \\
\mbox{subject to} &
x(0)=x_0,\;\;E x(T) = z, \;\;(x(\cdot),\pi(\cdot)) \; \mbox{ admissible.}
\end{array}
\end{equation}
Moreover, the problem is called {\it feasible} (with respect to $z$)
if there is at least
one admissible portfolio satisfying $E x(T) = z$.
An optimal portfolio to (\ref{equ007}),
if it ever exists, is called
an {\it efficient portfolio} with respect to $z$, and the corresponding
point on the diagram is called
an {\it efficient point}.
The set of all the efficient points (with different values of $z$)
is called the {\it efficient frontier}
(at $T$).}
\eDef

In the preceding definition, 
the parameter $z$ is restricted
to be no less than $x_0 e^{ \int_0^T r(t)dt}$,  the risk-free
terminal payoff.
Hence, as standard with the single-period case, we are
interested only in the {\it non-satiation} portion of the minimum-variance
set, or the upper portion of the left boundary of the admissible region.




\newsection{EFFICIENT PORTFOLIOS AND FRONTIER}{section3}{0.28in}
In this section, we highlight some existing results on the explicit
solution to the mean--variance portfolio selection problem (\ref{equ007}).

Denote the {\it risk premium function}
\bq{arbitrage} \theta(t)\equiv
(\theta_1(t),\cdots,\theta_m(t)):=B(t)'(\si(t)')^{-1}.
\end{equation}

The following result gives a complete solution to problem (\ref{equ007}).

\bthm{th1}
If $\sum_{i=1}^m\int_0^T|\mu_i(t)-r(t)|dt\neq 0$, then
problem (\ref{equ007})
is feasible for every $z\geq x_0 e^{ \int_0^T r(t)dt}$.
Moreover, the efficient
portfolio corresponding to each given
$z \geq x_0 e^{ \int_0^T r(t)dt}$ can be uniquely represented as
\bq{equ0112}
\pi^*(t)\equiv (\pi^*_1(t),\cdots,\pi^*_m(t))'
=-[\sigma(t)\sigma(t)']^{-1}B(t)[x^*(t)-\gamma e^{-\int_t^T
r(s)ds}],
\end{equation}
where $x^*(\cdot)$ is the corresponding wealth process and
  \bq{equ012}
    \gamma = \frac{z - x_0e^{\int_0^T [r(t)-|\theta(t)|^2]dt}} {1-e^{-\int_0^T
|\theta(t)|^2dt}}>0.
  \end{equation}
Moreover,
the corresponding minimum variance can be expressed as
  \bq{equ013}
    \mbox{{\rm Var} $x^*(T)$} = \frac{1}
      {e^{\int_0^T |\theta(t)|^2dt}- 1 }
      \left[z - x_0 e^{ \int_0^T r(t)dt}\right]^2, \;\;\; z\geq x_0 e^{
\int_0^T r(t)dt}.
  \end{equation}
\ethm

Equation (\ref{equ013}) gives
\bq{equ014}
z\equiv Ex^*(T) = x_0 e^{\int_0^T r(t)dt}
            + \sqrt{e^{\int_0^T |\theta(t)|^2dt}- 1}\sigma_{x^*(T)}.
\end{equation}

Define ${R}^*(t): = (x^*(t) - x_0)/x_0$, the return rate of
an efficient strategy 
at time $t$. Then by virtue of (\ref{equ013}) we immediately obtain the
efficient frontier.

\bthm{cor0}
The efficient frontier is
\bq{return}
E[R^*(T)]=R_f(T)+\sqrt{e^{\int_0^T |\theta(t)|^2dt}- 1 }
\sigma_{R^*(T)},
\end{equation}
where $R_f(T):=e^{\int_0^T r(t)dt} - 1$ is the risk-free return rate
over the entire horizon $[0,T]$.
\ethm

Hence, the efficient frontier is
a {\it straight line} which is the
continuous-time analog to the {\it capital market line}\footnote{The fact
that in the continuous-time setting the efficient frontier
is {\it still} a straight line may seem, at first sight, to be
a routine (and, shall we say, boring)
extension of the single-period case. However, as will be discussed
in the sequel, this fact is not to be taken lightly.}
 of the
classical single-period model [see, e.g., Sharpe (1964)].


\newsection{PREMIUM OF DYNAMIC TRADING}{s3}{0.28in}

As discussed in the introduction
the efficient frontier in the single-period case
is tangent to the risky region;
see Figure \ref{fig1a}.
Consequently, there is
a portfolio
consisting of only the risky assets, i.e., the tangent portfolio,
whose Sharpe ratio is the same as that of any efficient portfolio.
In particular, if there is only {\it one} risky asset (e.g., in a
Black--Scholes market), then
the Sharpe ratios of an efficient portfolio and the risky asset coincide.
However, this is no longer true
in the continuous-time setting, as shown in the following theorem.

\bthm{thm-bs}
In a Black--Scholes market where
there is one risk-free asset with a short rate $r>0$
and one risky asset with an appreciation rate $\mu>r$
and a volatility rate $\si>0$, the Sharpe ratio of any
mean--variance efficient portfolio is always strictly greater than
that of the risky asset.
\ethm

This result was first observed by Richardson (1989) via simulations, and
proved by Bajeux-Besnainous and Portait (1998), pp. 87--88. We will supply a proof
in appendix not only for the convenience of the reader, but -- more
importantly -- for understanding why the proof falls apart when there
are multiple risky assets and the risky region contains all the {\it continuously} traded (instead
of buy-and-hold) portfolios.

\medskip

Theorem \ref{thm-bs} is exemplified by the following example.

\beg{ex1}
{\rm Consider a
Black--Scholes market where there is one risky asset (the stock) with
$\mu = 0.12$ and $\sigma=0.15$, and a risk-free rate (continuously compounding)
$r = 0.06$. Assume an investment period $T=1$
(year). Then the risk-free return rate over one year is
$e^{0.06}-1=0.0618$.
According to (\ref{return}) the Sharpe ratio
of the efficient frontier
is 
\[ \sqrt{e^{\int_0^T |\theta(t)|^2dt}- 1 }=\sqrt{e^{(0.12-0.06)^2/0.15^2}-1}=0.4165. \]
On the other hand,
to determine the position of the stock on the
diagram we calculate the
return rate of the stock
\bq{mmean}
E[R(1)]=e^{\mu}-1=e^{0.12}-1=0.1275,
\end{equation}
and its standard deviation
\bq{msd}
    \sigma_{R(1)} =e^{\mu}\sqrt{e^{\sigma^2}-1}=e^{0.12}\sqrt{e^{0.15^2}-1}=  0.1701.
\end{equation}
Hence the Sharpe ratio of the stock is
\[ (0.1275-0.0618)/0.1701 = 0.3862 < 0.4165.\]
This means that the efficient frontier
lies much {above} the
stock. More precisely, the inclusion of the risk-free security
increases the Sharpe ratio by approximately 7.85\%.}
\eeg

Theorem \ref{thm-bs} demonstrates an intriguing difference between
the single-period (static) and continuous-time (dynamic) cases,
at least for the Black--Scholes market: the
efficient frontier is strictly separated from the risky region in
the latter case.
But the result is really not that surprising if one thinks a
little deeper: the phenomenon can be explained by the special
structure of the Black--Scholes market, in particular  the
availability of only {\it one} risky asset. The portfolio consisting of
the risky asset is essentially a buy-and-hold strategy owing to the lack
of another risky asset, thereby it is not taking advantage of
dynamic trading as enjoyed by an efficient portfolio which could
continuously adjust the weight between the risk-free and risky assets.
This is why the risky asset underperforms -- in terms of the Sharpe
ratio -- the efficient portfolio.

In the case of multiple risky assets, the same argument above applies
to yield that the efficient frontier is strictly separated from
the {\it buy-and-hold} risky region, as portrayed by Figure 1 in Bajeux-Besnainous and Portait (1998).

So, what if now we have multiple risky assets and the risky region
is generated by all the dynamically changing risky portfolios?
The dynamic risky region is much larger; would there be a chance that
it is large enough to touch the efficient frontier?

The answer to the last question is no: we are to establish the strict separation for a general continuous-time
market involving {\it multiple} stocks and
time-varying market parameters.
Note that we are no longer able to prove this using the direct calculation
as in the proof of Theorem \ref{thm-bs} (see appendix), because it is
not possible, at least for us, to obtain the expression for the (dynamic) risky
region.
Instead we will derive a
property of an efficient portfolio which {\it implies}
the said strict separation. In doing so we need other properties that are
interesting in their own rights.

\bthm{bdd}
Let $x^*(\cdot)$ be the wealth process under the efficient portfolio
corresponding to $z\geq x_0 e^{ \int_0^T r(t)dt}$. Then
\bq{bd1}
P\left\{x^*(t)\leq \gamma e^{-\int_t^T r(s)ds}\;\;\forall t\in[0,T]\right\}=1.
\end{equation}
Moreover, the inequality above is strict if and only if
$z>x_0 e^{ \int_0^T r(t)dt}$.
\ethm

This theorem implies that, 
with probability 1
the wealth under an efficient strategy must be capped at any time by
the present value of $\gamma$, which is
a {\it deterministic} constant depending only on the target $z$. In particular,
with probability 1 the terminal wealth will never exceed $\gamma$.
Notice such a property
is {unavailable} in the single-period case.

\bthm{cor1}
Let $\pi^*(\cdot)$ be an efficient portfolio corresponding to
$z>x_0 e^{ \int_0^T r(t)dt}$. If at $t_0\in[0,T]$, $\mu_i(t_0)\neq r(t_0)$
for some $i$, then
\bq{nostock}
P\left\{\pi^*(t_0)\neq 0\right\}=1. 
\end{equation}
\ethm

This result suggests that any efficient strategy (other than the
risk-free one) invests in at least one {risky} asset {whenever}
 any one of the risky appreciation rates is different from
the risk-free rate.

Notwithstanding the preceding result,
any efficient strategy must also invest in the risk-free asset
at {any} time, as
shown in the following theorem.

\bthm{always-risk-free}
Let $\pi^*(\cdot)$ be an efficient portfolio corresponding to
$z\geq x_0 e^{ \int_0^T r(t)dt}$.
Then we must have
\bq{nobond}
P\{\pi_0^*(t)\neq 0\}>0,\;\;\forall t\in(0,T],
\end{equation}
where $\pi_0^*(t)$ is the allocation to the bond
at $t$.
\ethm

{\red We knew that the efficient frontier was a straight line which by definition must lie above the risky hyperbola. Now, Theorem \ref{always-risk-free} excludes the possibility that
the former intersects with the latter (because any
efficient portfolio must invest in the risk-free asset at any
time). In other words,
the efficient frontier line is {\it strictly}
separated from the dynamic risky region, and hence must lie {\it strictly} above the risky hyperbola, as indicated by
Figure \ref{fig1b}.}



We can now formally state the following main result of this paper.

\bthm{premium}
{\it The Sharpe ratio of any continuous-time
mean--variance efficient portfolio is always strictly greater than
that of any admissible portfolio consisting of only the basic risky assets.}
\ethm

One
implication of Theorem \ref{premium}
is that the availability of {\it dynamic} trading
helps increase the Sharpe ratio of the efficient frontier compared with
static trading.
In the case of
Example \ref{ex1} (although this example is too simple to be a really good one),  where all the market data are quite typical,
the
Sharpe ratio of an efficient frontier is 0.4165 compared to
0.3862 of the stock -- an increase of nearly 8\% by dynamic trading. We
call this ``the premium of dynamic trading''.

Another more intriguing (albeit delicate) implication of Theorem \ref{premium}
is that the {\it availability} of a {risk-free} asset also helps
increase the Sharpe ratio of one's portfolios in the continuous-time
setting. This, again, is in
sharp contrast with the single-period case. To elaborate,
consider first the single-period case. If there is no
risk-free asset available, then the portfolio that produces the {\it highest}
Sharpe ratio is the tangent portfolio. Hence, the availability of
an additional risk-free asset does not yield a Sharpe ratio higher than
that of the tangent one; see Figure 1. However, in the continuous-time setting, the
newly discovered strict separation of the efficient frontier from the
risky hyperbola suggests that a higher Shape ratio is achieved when the risk-free
asset is available for trading. 

\begin{figure}[h]
\begin{center}
\includegraphics[scale=0.65]{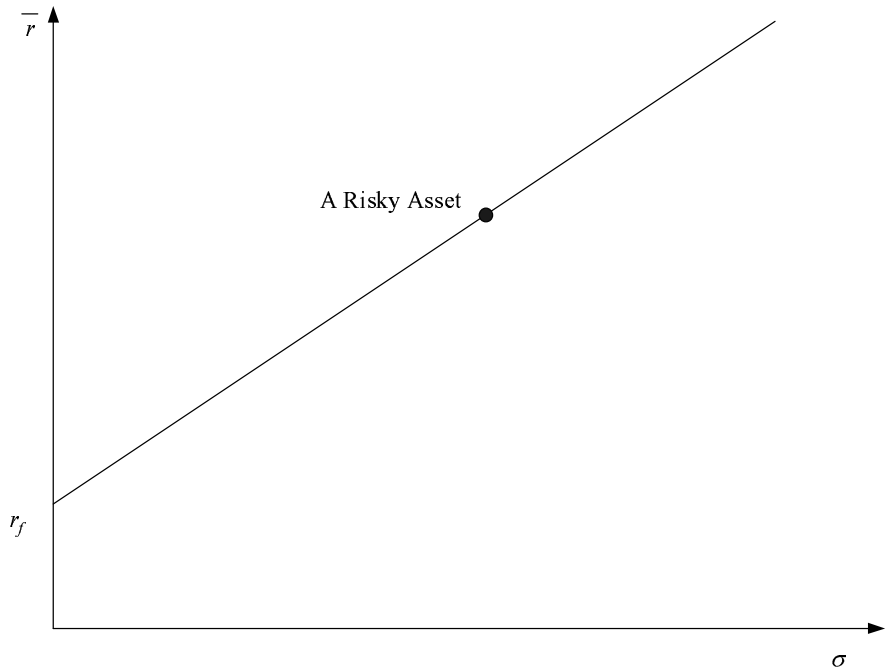}
\caption{Combination of the Risk-Free Asset and a Risky One:
Single-Period Case}\label{fig2a}
\end{center}
\end{figure}

\newpage

\newsection{CAUSE OF THE PREMIUM}{s5}{0.28in}

\begin{figure}[h]
\begin{center}
\includegraphics[scale=0.65]{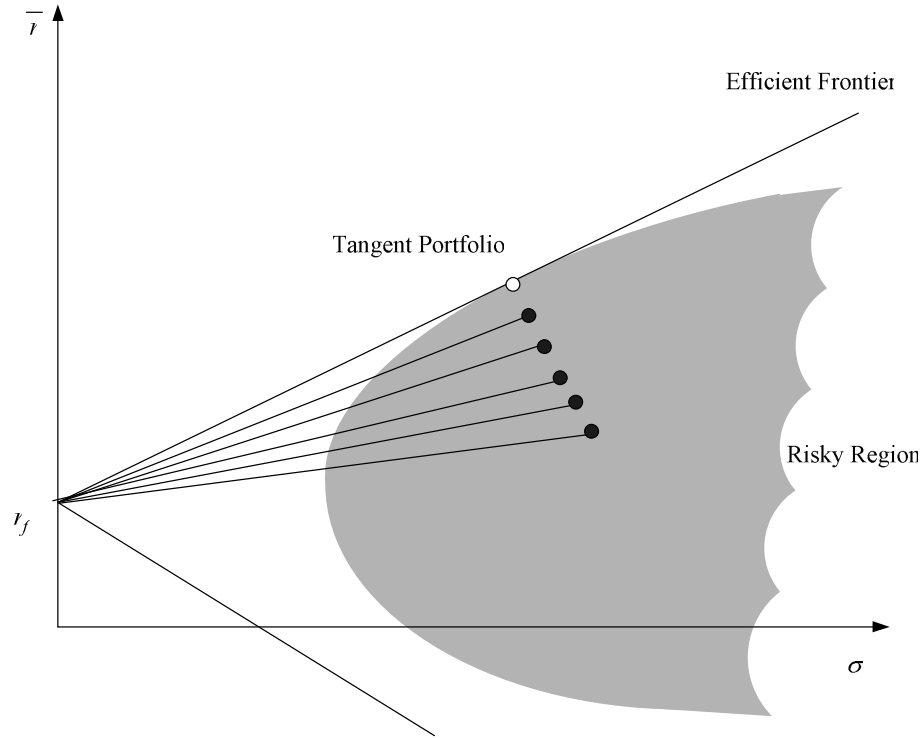}
\caption{Admissible Region: Single-Period Case}\label{fig2b}
\end{center}
\end{figure}

But what is the {\it fundamental} reason underlying such a great difference
between the continuous-time case and the single-period one?
This is best explained\footnote{The discussions in this section
apply, mutatis mutandis, also to a dynamic, discrete-time model.}
 by first recalling the
reason for the tangent line in the single-period case.
Consider the risk-free asset and any risky asset, which are two
points on the diagram. These two assets
are combined to form a portfolio using a weight of $\alpha$ for the
risk-free asset, where $\alpha\in \Re$. It is easy to show [see,
e.g., Luenberger (1998), p. 165] that all such portfolios are on the
straight line crossing the original two assets (see Figure
\ref{fig2a}). For {\it each} risky asset there is such a straight line; thus
the admissible region (the one including both
risk-free and risky assets) is a triangularly shaped region with
the upper boundary being the straight line tangent to the risky hyperbola\footnote{This also suggests that
the efficient frontier {\it should} be a straight line.}; see Figure \ref{fig2b}.

\begin{figure}[h]
\begin{center}
\includegraphics[scale=0.65]{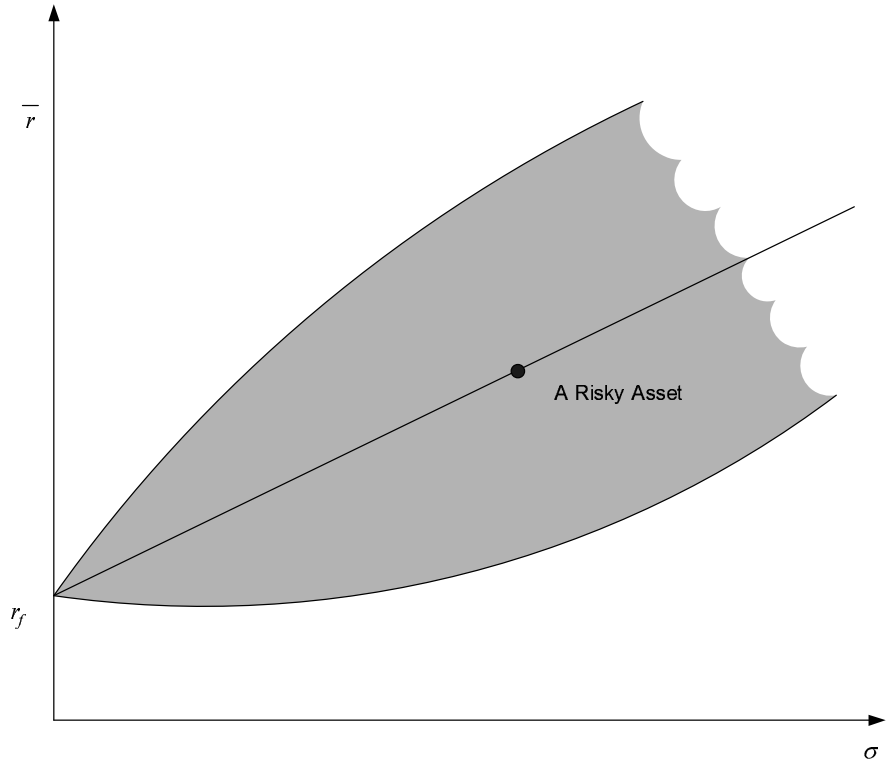}
\caption{Combination of the Risk-Free Asset and a Risky One:
Continuous-Time Case}\label{fig3a}
\end{center}
\end{figure}


Now, in the continuous-time case, let us also start with two
assets, a risk-free asset represented by the portfolio
$\pi_f(\cdot)\equiv 0$, and a risky asset represented by
$\pi(\cdot)\neq \pi_f(\cdot)$. We use these two assets to form new
portfolios of the form
$\pi_\alpha(t)=[1-\alpha(t)]\pi_f(t)+\alpha(t)\pi(t)\equiv \alpha(t)\pi(t)$, where
$\alpha(\cdot)$ is any progressively measurable process so long as
$\pi_\alpha(\cdot)$ is admissible. Let $x(\cdot)$ be the wealth process corresponding
to $\pi(\cdot)$.
If $\alpha(t)\equiv
\alpha\in\Re$; that is, the corresponding $\pi_\alpha(\cdot)$ is a
{\it buy and hold} combination of the two,
then it is immediate from the linearity of the wealth equation
(\ref{system2}) that $\alpha x(\cdot)$ is the wealth process corresponding
to $\pi_\alpha(t)$ whose terminal $(\sigma,\bar r)$-value
lies on the
straight line connecting those of the two original assets. However, in
general $\alpha(\cdot)$ can be any appropriate {\it stochastic
process} (in other words, the new portfolio $\pi_\alpha(\cdot)$
is in general a {\it dynamic} combination of the two assets);
so $\alpha(\cdot)x(\cdot)$ is no longer necessarily the wealth process under
$\pi_\alpha(\cdot)$ or it may not lie on the straight line connecting the two original assets. As a consequence, all the new
portfolios (with all the possible processes $\alpha(\cdot)$)
forming from the original two assets
generate a much larger {\it region} as indicated in Figure
\ref{fig3a}.

\begin{figure}[h]
\begin{center}
\includegraphics[scale=0.65]{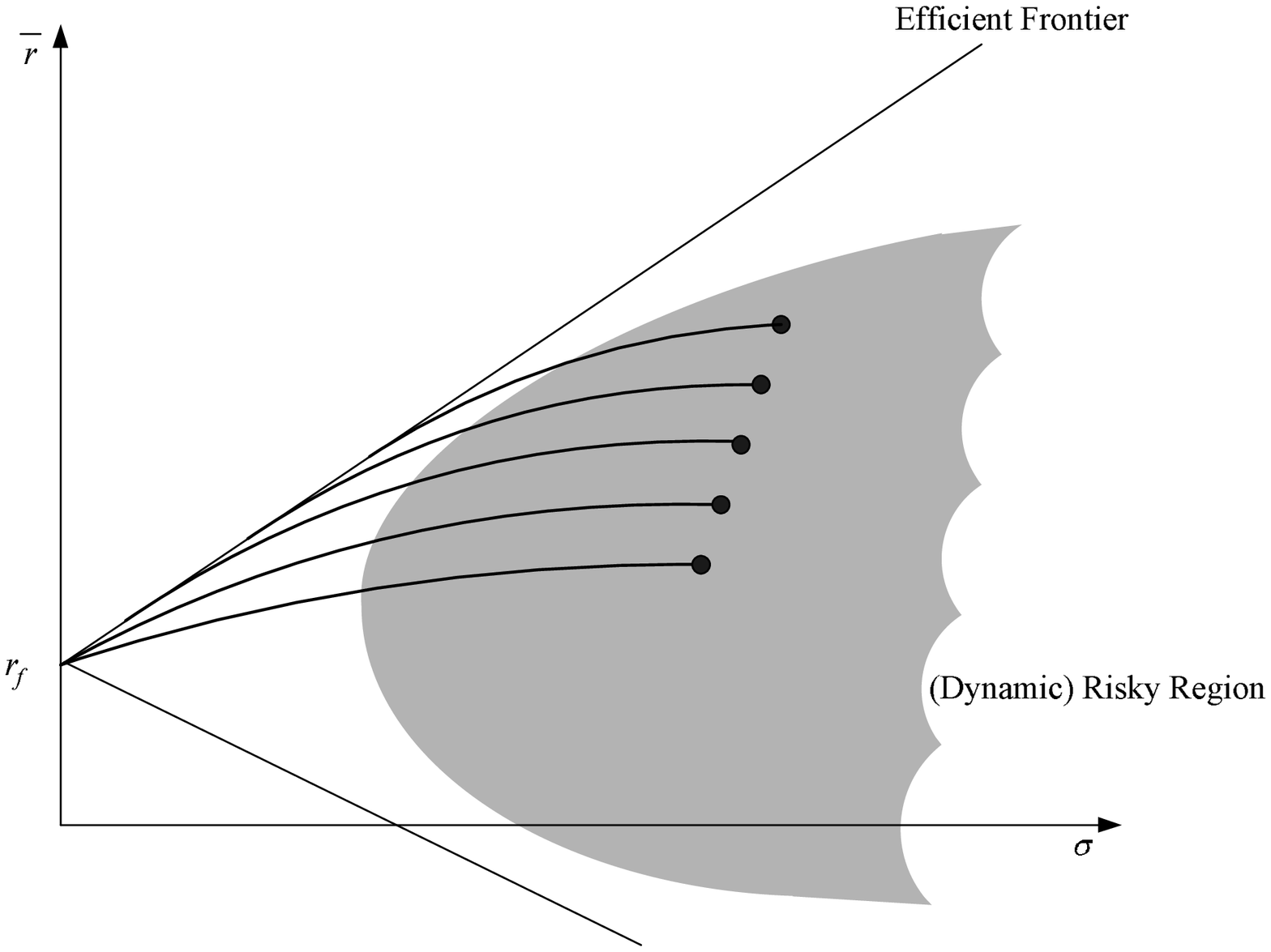}
\caption{Admissible Region: Continuous-Time Case}\label{fig3b}
\end{center}
\end{figure}

To discuss on the shape of the admissible region in the continuous-time
case (to be precise we are only interested in the {\it upper} left
boundary of the admissible region), we first construct the dynamic risky region
defined by the risky assets only, which is the shaded
region in Figure \ref{fig3b}. Next, for {\it each} portfolio in this
region we make combinations with the risk-free asset. As discussed
above these new portfolios form a solid two-dimensional region
with its upper left boundary being a curved line in general
(which could be
a straightline in some special cases -- such as the Black--Scholes case). There is such a
curved line corresponding to every asset in the risky
region. The {\it envelope} of these lines forms the upper left boundary of the
entire admissible region (containing both the risk-free and risky assets), which
is the efficient frontier we are seeking. It is these
curved lines that push the efficient frontier away from the
risky region; and hence the surprising phenomenon
stipulated in Theorem \ref{always-risk-free}.


Last but not least, the fact that in continuous time the mean--variance
efficient frontier is still a straight line, derived in
Bajeux-Besnainous and Portait (1998) and
Zhou and Li (2000), is no longer a mere routine
extension of its single-period counterpart, and should not be taken lightly.
In fact, the efficient frontier is the envelope of infinitely many {\it curved}
lines, which {\it turns out} to be a straight line. This is quite an unusual
coincidence, indeed.

\newsection{CONCLUDING REMARKS}{s6}{0.28in}

This paper disclosed some rather unexpected phenomena associated with a continuous-time
mean--variance market, suggesting
that continuous-time financial models have more
complex, sometimes strikingly different, structures and properties than its single-period
counterpart.
One should appreciate that continuous-time is probably
a closer representation of the real-world investment today than its discrete-time
counterpart, not to mention its analytical tractability which enables us to
elicit important economic insights from results that are often explicit. On the
other hand, in the realm of continuous-time asset allocation and asset
pricing the
literature has been dominated by the expected utility maximization (EUM) models. However, ``few if any agents know their utility functions; nor do the functions which financial engineers and financial economists find analytically convenient necessarily represent a particular agent's
attitude towards risk and return'' [Markowitz (2004)]. It is fair to say
mean--variance (including the related Sharpe ratio) remains nowadays one of the most commonly used measures to assess the
performance of fund managers. Theoretically, while mean--variance
has some inherent drawbacks such as the sensitivity of the final solutions on
the investment opportunity set, and the inconsistency
with the dynamic programming principle in the continuous-time setting, recent
studies have revealed some redeeming quality of continuous-time mean--variance
efficient policies. For example, it is shown in Li and Zhou (2006) that,
under the same setting as in this paper,
a mean--variance efficient portfolio realizes the (discounted)
targeted return on or before the terminal date with a probability greater than
$0.8072$. This number is universal irrespective of the
opportunity set, the
targeted return, and the investment horizon.
This, together with the new findings in this paper, suggests that
it is necessary and important to revisit mean--variance models, in terms of
both asset allocation and asset pricing, for dynamic markets.

In this paper there are several assumptions on the underlying model, such as free of
transaction cost, market completeness, and deterministic investment opportunity set.
Mean--variance models with proportional transaction costs have been recently studied
by Dai, Xu and Zhou (2008). It would therefore be interesting to extend the results of this paper to the case with  transaction costs.
On the other hand, we believe our results can be
extended to the incomplete market case, noting the
results on mean--variance model in an incomplete market by Lim (2004).
The case with {\it stochastic} investment opportunity, however,  is still open,
as Theorems \ref{bdd} -- \ref{always-risk-free} may no longer hold true.

\vspace{17mm}

\newpage

\noindent{\Large\bf Appendix: Proofs}

\appendix

\bigskip

{\sc Proof of Theorem \ref{th1}:}
The assertion about the
feasibility of the problem is an immediate consequence of Lim and
Zhou (2002), Corollary 5.1.
The efficient feedback strategies (\ref{equ0112}) and the 
expression (\ref{equ013}) are derived in Zhou and Li (2000),
eq. (5.12) and Thm. 6.1.
Finally, that $\gamma>0$ is seen from the facts that $x_0>0$,
$z\geq x_0 e^{ \int_0^T r(t)dt}$ and $|\theta(t)|>0$ (due to $B(t)\neq 0$)
$\mbox{{\rm a.e.} }t\in [0,T]$.
\eof

{\sc Proof of Theorem \ref{thm-bs}:} To prove this theorem we need the
following technical lemma.

\blemma{pro-bs}
  For any $b > 0$
  \bq{eq-bs1}
    (e^{bx}-1)(e^{\frac{b}{x}}-1)\geq (e^b-1)^2\;\;\; \forall x>0.
  \end{equation}
\elemma

{\sc Proof:} The case when $x=1$ is trivial. Hence due to the symmetry we need only to show
(\ref{eq-bs1}) for any $0<x<1$.

Let $f(x) = (e^{bx}-1)(e^{\frac{b}{x}}-1)$, $0<x\leq 1$. Then
 \[
f'(x) = be^{\frac{b}{x}+bx} \left(1  -\frac{1}{x^2} +\frac{1}{x^2}e^{-bx} - e^{{\red -}\frac{b}{x}}\right).
 \]
Denote $g(x) = 1  -\frac{1}{x^2} +\frac{1}{x^2}e^{-bx} -
e^{-\frac{b}{x}}$, $0<x\leq 1$. We have
\[
  \begin{array}{llll}
  g'(x)& = & \frac{1}{x^3} (2  - 2e^{-bx} - bxe^{-bx} -
  bxe^{-\frac{b}{x}})&\\
  & > & \frac{2}{x^3} \left[1  - (1+bx)e^{-bx}\right] & 
\\
  & > & 0, & \forall 0<x<1.
  \end{array}
 \]
So $g(x)< g(1)=0$ $\forall 0<x<1$, or
$f'(x)<0$ $\forall 0<x<1$. This leads to $f(x)\geq f(1)=(e^b-1)^2$.
\eof

We now prove Theorem \ref{thm-bs},  which is to show that
 \bq{eq-bs2}
   \frac{E[R^*(T)] - R_f(T)}{\si_{R^*(T)}} >  \frac{E[R(T)] -
   R_f(T)}{\si_{R(T)}} ,
 \end{equation}
 where $R^*(T)$ is the rate of return of any efficient portfolio,
 $R_f(T)$ is the risk free rate of return,  and $R(T)$
is the rate of return of the risky asset, over $[0,T]$.
Note $E[R(T)]=e^{\mu T}-1$, $\si_{R(T)}^2=e^{(2\mu + \si^2)T}-e^{2\mu T}$, and
$R_f(T)=e^{rT}-1$. So
according to \refs{return}, \refs{eq-bs2} is equivalent to
\bq{eq-bs22}
e^{(\frac{\mu-r}{\si})^2T} - 1>\frac{(e^{\mu T}-e^{rT})^2}{e^{(2\mu + \si^2)T}-e^{2\mu T}}.
\end{equation}
Now,
 \[
e^{(\frac{\mu-r}{\si})^2T} - 1 - \frac{(e^{\mu T}-e^{rT})^2}{e^{(2\mu + \si^2)T}-e^{2\mu
   T}}
   =  \frac{e^{2\mu T}(e^{\si^2 T}-1)\left(e^{((\mu-r)^2/\si^2) T}-1\right)-e^{2rT}(e^{(\mu-r)T}-1)^2}{e^{(2\mu + \si^2)T}-e^{2\mu
   T}}.
 \]
 Considering the numerator of the above and noting $\mu>r$, we have
  \[
   \begin{array}{lll}
   & e^{2\mu T}(e^{\si^2 T}-1)(e^{((\mu-r)^2/\si^2)
   T}-1)-e^{2rT}(e^{(\mu-r)T}-1)^2\\
   > & e^{2\mu T} [(e^{\si^2 T}-1)(e^{((\mu-r)^2/\si^2)
   T}-1)-(e^{(\mu-r)T}-1)^2] &\\ 
   > & (e^{\si^2 T}-1)(e^{((\mu-r)^2/\si^2)
   T}-1)-(e^{(\mu-r)T}-1)^2. & 
   \end{array}
  \]
Letting $(\mu - r)T = b$ and $\si^2 =(\mu - r) x$ where $x>0$,
we have
  \[
   \begin{array}{lll}
   & (e^{\si^2 T}-1)(e^{((\mu-r)^2/\si^2)
   T}-1)-(e^{(\mu-r)T}-1)^2 & \\ 
   = & (e^{bx}-1)(e^{b/x}-1)-(e^{b}-1)^2 \\
   \geq & 0 & 
   \end{array}
  \]
by virtue of Lemma \ref{pro-bs}. The proof is complete.
\eof

{\sc Proof of Theorem \ref{bdd}:}
Set $y(t):=x^*(t)-\gamma e^{-\int_t^Tr(s)ds}$. Using
the wealth equation {\red (\ref{system2})} that $x^*(\cdot)$ satisfies and {\red
applying (\ref{equ0112})}, we deduce
\[\left\{\begin{array}{l}
dy(t)=[r(t)-|\theta(t)|^2]y(t)dt-\theta(t)y(t)dW(t),\\
y(0)=\frac{x_0-ze^{-\int_0^T r(t)dt}}{1-e^{-\int_0^T|\theta(t)|^2dt}}\leq 0.
\end{array}\right.
\]
The above equation has a unique solution
\[ y(t)=y(0)\exp\left\{\int_0^t[r(s)-\frac{3}{2}|\theta(s)|^2]ds-\int_0^t\theta(s)dW(s)\right\}\leq 0,
\]
and the inequality is strict if and only if $y(0)<0$, or $z>x_0 e^{ \int_0^T r(t)dt}$.
\eof

{\sc Proof of Theorem \ref{cor1}:}
Since $x^*(t)\neq \gamma e^{-\int_t^T r(s)ds}$, the
result follows directly from Theorem \ref{bdd} and (\ref{equ0112}).
\eof

{\sc Proof of Theorem \ref{always-risk-free}:}
If $z=x_0 e^{ \int_0^T r(t)dt}$, then the theorem holds trivially.
So we assume that $z>x_0 e^{ \int_0^T r(t)dt}$.
According to Theorem
\ref{th1}, $\pi^*(\cdot)$ can be written as
\bq{eff_mar}
\pi^*(t)  \equiv (\pi_1^*(t), \cdots, \pi_m^*(t))'
 =-[\sigma(t)\sigma(t)']^{-1}B(t)[x^*(t)-\gamma e^{-\int_t^Tr(s)ds}],
\end{equation}
for {\it some} $\gamma>0$.
If (\ref{nobond}) is not true, then there is $\bar t\in(0,T]$ such that
$P\{\pi_0^*(\bar t)=0\}=1$, or
{\red \bq{nobond1}
P\{x^*(\bar t)=\sum_{i=1}^m\pi_i^*(\bar t)\}=1.
\end{equation}}
Summing all the components of $\pi^*(t)$ in (\ref{eff_mar}) at $t=\bar t$,
we have
\bq{keyeq}
x^*(\bar t)= \sum_{i=1}^m \pi_i^*(\bar t)
 = -\eta(\bar t) [x^*(\bar t)-\gamma e^{-\int_{\bar t}^Tr(s)ds}],
\end{equation}
where $\eta(t): = {\bf e}' [\sigma(t)\sigma(t)']^{-1}B(t)$ with
${\bf e}= (1, 1, \cdots, 1)'$. Note that $\eta(t)\neq -1\;\forall
t\in[0,T)$, for otherwise (\ref{keyeq}) would yield $\gamma=0$,
contradicting (\ref{equ012}). Hence, it follows from (\ref{keyeq})
that \bq{eff_mar_x}
  x^*(\bar t) = \frac{\gamma \eta(\bar t) e^{-\int_{\bar t}^T r(s) ds}}{1 + \eta(\bar t)}.
\end{equation}
In other words,
the wealth at time $\bar t$ is a {\it deterministic} quantity.

However, the wealth equation {\red (\ref{system2})} that $x^*(\cdot)$ satisfies up to the time $\bar t$
can be rewritten as
\bq{system3}
\left\{\begin{array}{l}
dx^*(t)=[r(t)x^*(t)+\theta(t)(\si(t)'\pi^*(t))]dt+(\si(t)'\pi^*(t))'dW(t),
\;\;t\in[0,\bar t),\\
x^*(\bar t) = \frac{\gamma \eta(\bar t) e^{-\int_{\bar t}^T r(s) ds}}{1 + \eta(\bar t)}.
\end{array}\right.
\end{equation}
The above is a linear backward stochastic differential equation (BSDE)
with deterministic
linear coefficients as well as a deterministic terminal condition.
Hence by the uniqueness of its solution 
we must have
$\si(t)'\pi^*(t)=0\;\; \as,\; \mbox{{\rm a.e.}}\; t \in [0,\bar t]$.
Appealing to (\ref{non-degeneracy}), we conclude that
$\pi^*(t)={0}$, \as, $\mbox{{\rm a.e.}}\; t \in [0,\bar t].$
This
in turn implies $B(t)=0$ $\forall t\in[0,\bar t]$, in view of (\ref{eff_mar}), Theorem \ref{bdd},
and the fact that $B(t)$ is continuous in $t$.
So $\eta(t)=0$, $\forall t \in [0,\bar t]$, and hence $x^*(\bar t)=0$.
Again by the uniqueness of solution to BSDE (\ref{system3}), $x_0=0$, which is a contradiction.
\eof

\newpage

\begin{center}REFERENCES \end{center}

\singlespacing

\begin{description}

\item {\sc Bajeux-Besnainous, L., and R. Portait} (1998):
``Dynamic Asset Allocation in a Mean---Variance Framework,''
\emph{Management Sciences}, 44, 79--95.

\item {\sc Dai, M., Z. Xu, and X.Y. Zhou} (2008): ``Continuous-Time Mean--Variance Portfolio
Selection with Proportional
Transaction Costs,'' Working paper, University of Oxford.












\item \small K\scriptsize ARATZAS, \small I., \scriptsize AND \small
S.E. S\scriptsize HREVE \small (1999):
\emph{Methods of Mathematical Finance}.
New York: Springer-Verlag.

\item {\sc Li, D.,  and W.L. Ng} (2000):
``Optimal Dynamic Portfolio Selection: Multi-period
Mean-Variance Formulation,''
\emph{Mathematical Finance, 10}, 387--406.



\item {\sc Lim, A.E.B.} (2004): ``Quadratic Hedging and Mean--Variance Portfolio
Selection with Random Parameters in an Incomplete Market,'' \emph{Mathematics
of Operations Research}, {29}, 132--161.

\item \small L\scriptsize IM, \small A.E.B., \scriptsize AND \small
X.Y. Z\scriptsize HOU \small (2002):
``Mean-Variance Portfolio Selection with Random Parameters in a
Complete Market,"
\emph{Mathematics of Operations Research,} 27, 101--120.


\item {\sc Li, X. and X.Y. Zhou} (2006): ``Continuous-Time Mean--Variance
Efficiency: The 80\% Rule,'' {\it Annals of Applied Probability}, 16, 1751--1763.

\item \small L\scriptsize INTNER, \small J. (1965): ``The Valuation of
Risk Assets and The Selection
of Risky Investments in Stock Portfolios and Capital Budgets,"
\emph{Review of Economics and Statistics,} 47, 13--37.

\item {\sc Luenberger D.G.} (1998): \emph{Investment Science},
Oxford University Press, New York.

\item \small M\scriptsize ARKOWITZ, \small H. (1952): ``Portfolio
Selection,"
\emph{Journal of Finance,} 7, 77--91.



\item
-------- (1987):
\emph{Mean--Variance Analysis in Portfolio Choice
and Capital Markets}. New Hope, Pennsylvania: Frank J. Fabozzi Associates.

\item \small M\scriptsize ARKOWITZ, \small H., \scriptsize AND \small
X.Y. Z\scriptsize HOU \small (2004): {Private communication}.



\item \small M\scriptsize ERTON, \small R.C. (1972): ``An Analytic
Derivation of the Efficient
frontier," \emph{Journal of Financial and Quantitative Analysis,} 7, 1851--1872.


\item \small M\scriptsize OSSIN, \small J. (1966): ``Equilibrium in a
Capital Asset Market,"
\emph{Econometrica,} 34, 768--783.


\item {\sc Richardson, H.R.} (1989):
``A Minimum Variance Result in Continuous Trading Portfolio Optimization,''
\emph{Management Sciences}, 25, 1045--1055.



\item \small S\scriptsize HARPE, \small W.F. (1964): ``Capital Asset
Prices: A Theory of Market
Equilibrium under Conditions of Risk,"
\emph{Journal of Finance,} 19, 425--442.


\item \small T\scriptsize OBIN, \small J. (1958): ``Liquidity
Preference as Behavior Towards Risk,"
\emph{The Review of Economic Studies,} 25, 65--86.




\item
{\sc Zhou, X.Y., and D. Li}
 \small (2000): ``Continuous-Time Mean-Variance
Portfolio
Selection: A Stochastic LQ Framework," \emph{Applied Mathematics and
Optimization,} 42, 19--33.

\end{description}


\end{document}